\documentclass[twocolumn]{aastex63}
\usepackage{times}
\usepackage{amsmath}
\usepackage{textcomp}
\usepackage{amssymb}
\usepackage{natbib}
\usepackage{float}
\usepackage{color}
\usepackage{graphicx}

\newcommand{\msun}{M$_{\odot}$}
\newcommand{\lsun}{L$_{\odot}$}
\newcommand{\lum}{erg\,s$^{-1}$}

\newcommand{\phflux}{\mbox{${\rm \, ph \,\, cm^{-2} \, s^{-1}}$}}

\newcommand{\gm}{$\gamma$}

\received{\today}
\revised{\today}
\accepted{\today}
\submitjournal{ApJ}

\shorttitle{Gamma rays from the Ring and Bullet}
\shortauthors{Paliya and Saikia}

\begin{document}
\title{A Gamma-ray Emitting Collisional Ring Galaxy System in our Galactic Neighborhood}

\correspondingauthor{Vaidehi S. Paliya}
\email{vaidehi.s.paliya@gmail.com}

\author[0000-0001-7774-5308]{Vaidehi S. Paliya}
\affiliation{Inter-University Centre for Astronomy and Astrophysics (IUCAA), SPPU Campus, Pune 411007, India}
\author[0000-0002-4464-8023]{D. J. Saikia}
\affiliation{Inter-University Centre for Astronomy and Astrophysics (IUCAA), SPPU Campus, Pune 411007, India}

\begin{abstract}
The astrophysical \gm-ray photons carry the signatures of the violent phenomena happening on various astronomical scales in our Universe. This includes supernova remnants, pulsars, and pulsar wind nebulae in the Galactic environment and extragalactic relativistic jets associated with active galactic nuclei (AGN). However, $\sim$30\% of the \gm-ray sources detected with the Fermi Large Area Telescope lack multiwavelength counterpart association, precluding us from characterizing their origin. Here we report, for the first time, the association of a collisional ring galaxy system in our Galactic neighborhood (distance $\sim$10 Mpc), formed as a consequence of a smaller `bullet' galaxy piercing through a larger galaxy, as the multi-frequency counterpart of an unassociated \gm-ray source 4FGL~J1647.5$-$5724. The system, also known as ``Kathryn's Wheel", contains two dwarf irregular galaxies and an edge-on, late-type, spiral galaxy surrounded by a ring of star-forming knots. We utilized observations taken from the Neil Gehrels Swift observatory, Rapid ASKAP Continuum Survey, SuperCOSMOS H$\alpha$ Survey, Dark Energy Survey, and VIsible MultiObject Spectrograph at Very Large Telescope to ascertain the association with 4FGL~J1647.5$-$5724 and to explore the connection between the star-forming activities and the observed \gm-ray emission. We found that star-formation alone cannot explain the observed \gm-ray emission, and additional contribution likely from the pulsars/supernova remnants or buried AGN is required. We conclude that arcsecond/sub-arcsecond-scale observations of this extraordinary \gm-ray emitting galaxy collision will be needed to resolve the environment and explore the origin of cosmic rays.
 
\end{abstract}


\section{Introduction}
The identification of \gm-ray emitting astrophysical objects is one of the crucial research problems in high-energy astrophysics \citep[cf.][]{2020ApJS..248...23D}. Finding the source of \gm-ray emission is not only critical to understanding the radiative processes and the interaction of the particle population with the ambient environment but also to reveal the origin of the diffuse extragalactic \gm-ray background \citep[e.g.,][]{2013ApJ...768...54B,2014ApJ...780..161D,2016PhRvL.116o1105A,2020ApJ...894...88A,2022ApJ...933..213D}. Furthermore, with the advent of multi-messenger astronomy, the \gm-ray source population has emerged as a promising candidate for the cosmic neutrinos detected with the IceCube observatory \citep[cf.][]{2019ApJ...880..103G,2022Sci...378..538I}. In the extragalactic \gm-ray sky, blazars, i.e., radio-loud active galactic nuclei (AGN) hosting relativistic jets closely aligned with the line of sight, are the most abundant source class. On the other hand, radio galaxies and star-forming galaxies form a minor fraction ($<$10\%) of the known \gm-ray source population \citep[][]{2020ApJ...894...88A,2022ApJS..263...24A}. Despite several multi-frequency efforts, the counterparts of about 30\% of the \gm-ray sources are yet to be identified/associated \citep[see, e.g.,][]{2022ApJS..260...53A}. A significant fraction of the unassociated Fermi-Large Area Telescope (LAT) detected sources are expected to be jetted AGN \citep[e.g.,][]{2019ApJ...871...94K,2020A&A...638A.128M,2021ApJ...923...75K}. However, thanks to the sensitive, high-resolution multiwavelength data from the latest surveys, e.g., e-ROSITA, the discovery potential to identify a \gm-ray emitting, non-jetted source population remains high. 

4FGL~J1647.5$-$5724 is an unassociated \gm-ray source that first appeared in the second data release of the fourth catalog of the Fermi-LAT detected \gm-ray sources \citep[4FGL-DR2;][]{2020ApJS..247...33A}. In this Letter, we report the association of the collisional ring galaxy system {\it ``Kathryn's Wheel"} \citep[][]{2015MNRAS.452.3759P} as the low-frequency counterpart of 4FGL~J1647.5$-$5724. Such an enigmatic structure forms when a smaller `bullet' galaxy pierces through the minor axis of another larger galaxy close to its center \citep[see, e.g.,][]{1996FCPh...16..111A}. The produced shock wave sweeps up and kicks the gas out of the system, leaving behind a ring of star-forming regions and a gas-poor galaxy. A few examples of such systems are the Cartwheel galaxy, Arp 147, and Arp 148 \citep[][]{1964ApJ...140.1617B,1977MNRAS.178..473F,1992ApJ...399L..51G}. The \gm-ray emission in this collisional ring galaxy system is likely to originate from vigorous star-forming activities. However, additional contributions, such as from supernova remnants, pulsars, pulsar-wind nebulae, and buried AGN, are not ruled out. Since the \gm-ray emission detected from star-forming galaxies is thought to be produced by the interaction of cosmic rays with the dense interstellar medium \citep[e.g.,][]{1996ApJ...460..295P,2011ApJ...734..107L}, this system provides a unique opportunity to explore the sites of cosmic-ray acceleration and propagation due to its proximity to the Milky Way.

We adopt a flat cosmology with $H_0 = 70~{\rm km~s^{-1}~Mpc^{-1}}$ and $\Omega_{\rm M} = 0.3$.

\begin{figure*}
\hbox{\hspace{2.cm}
    \includegraphics[scale=0.33]{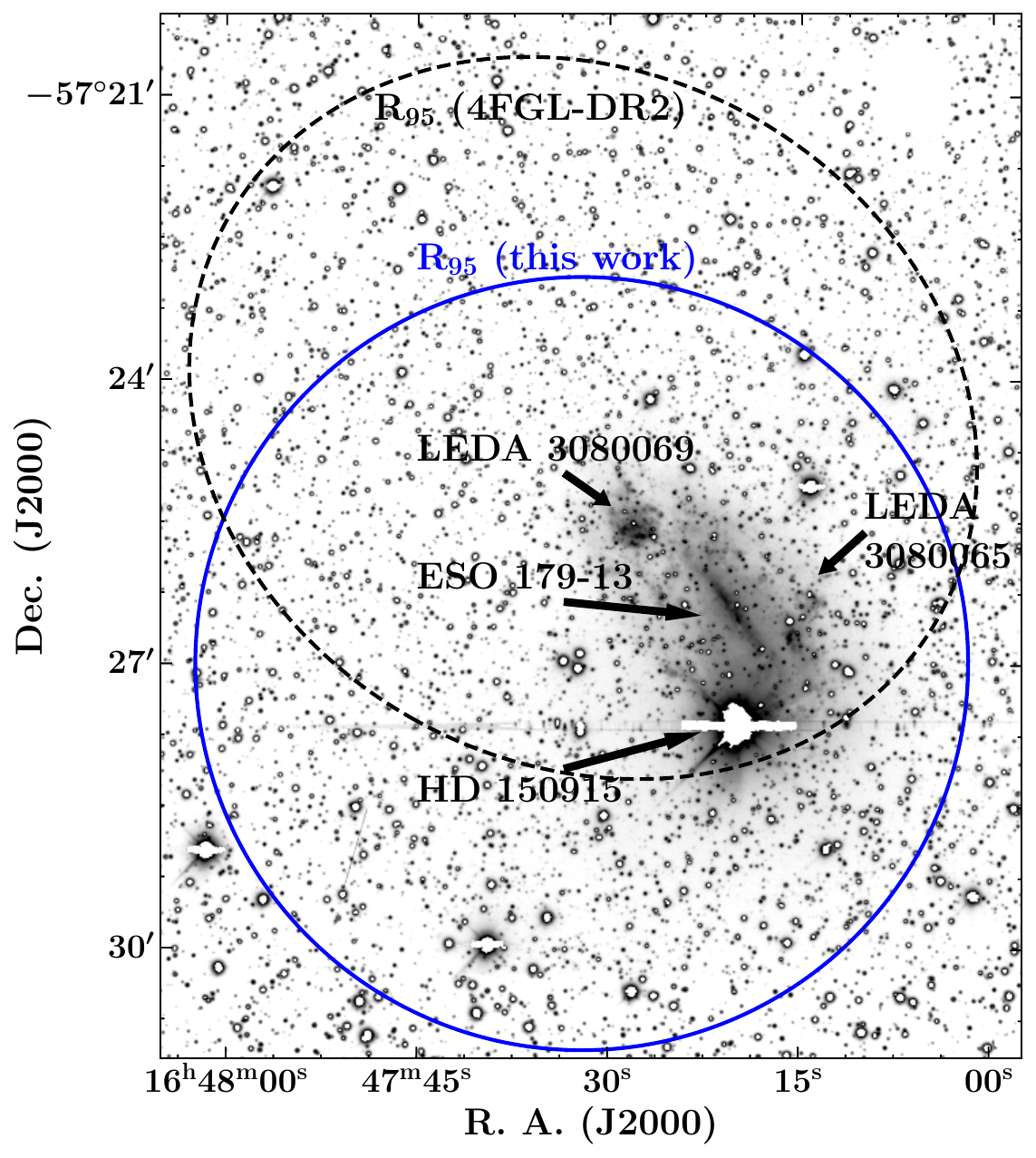}
    \includegraphics[scale=0.33]{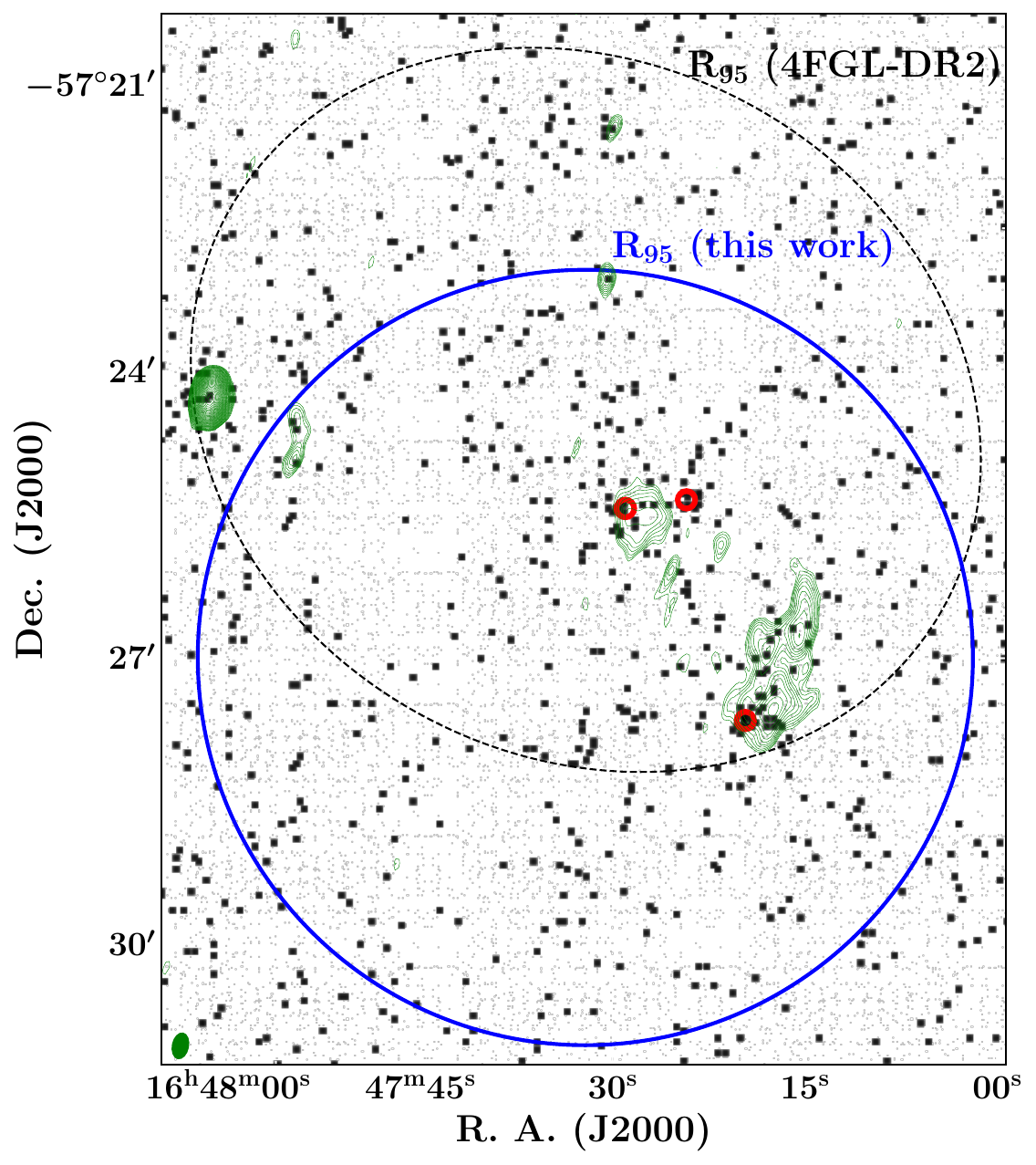}
    }
    \hbox{\hspace{4.5cm}
    \includegraphics[scale=0.45]{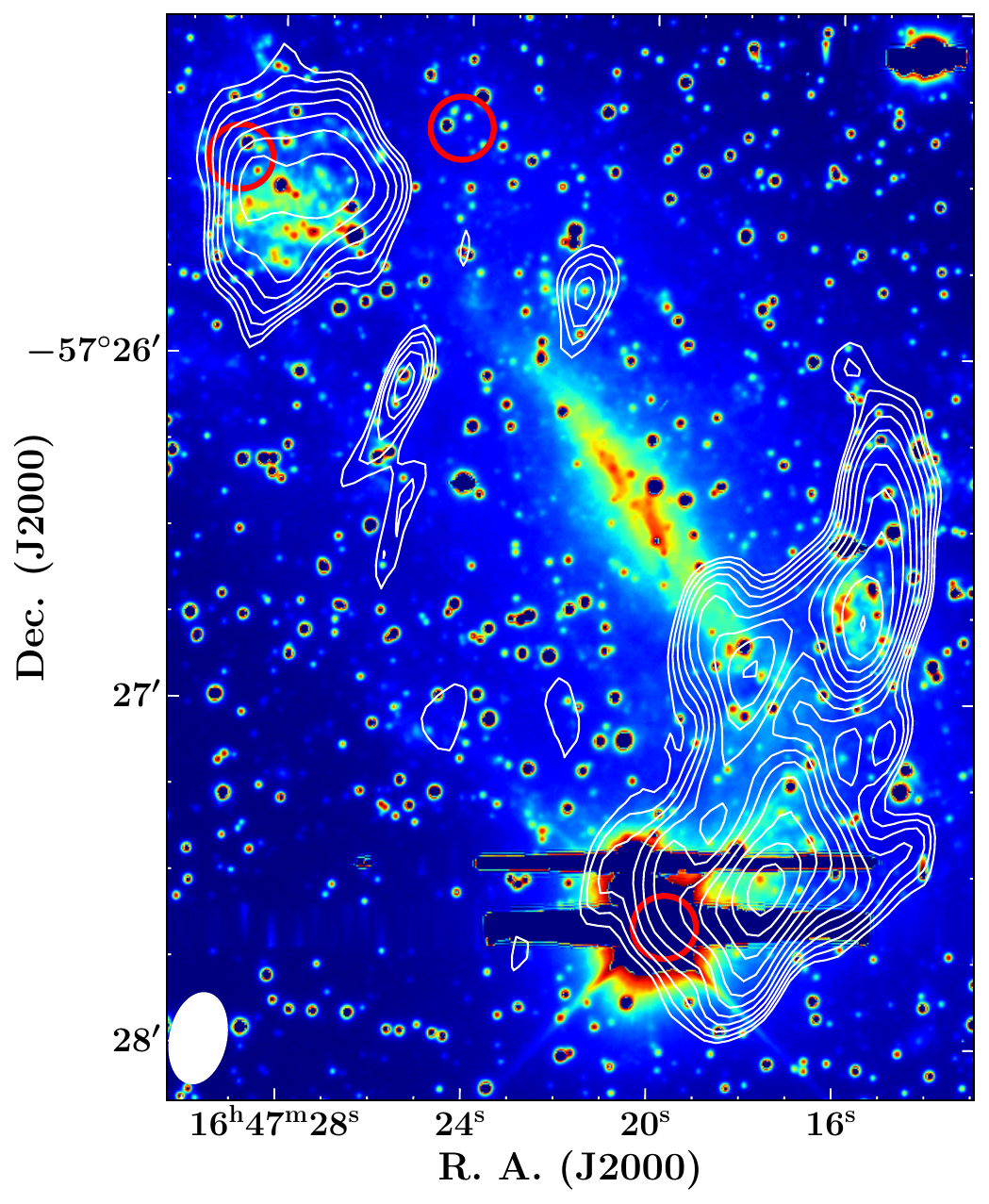}
    }
\caption{Top left: DECAM $g$-band image of {\it ``Kathryn's Wheel"}. The brightest pixels were masked, and the color scale was adjusted to enhance the faint features. Top right: Swift-XRT counts map of the same region. The red circles represent the identified X-ray sources and have a radius of 5$^{\prime\prime}$.6, estimated as the positional uncertainty in the optimized X-ray positions. The green contours refer to the 1.36 GHz emission observed by RACS. The contours start at three$\times$rms and increase in the multiple of $\sqrt{1.5}$. Bottom: Same as the top panels but zooming in on the system. Here, we have used DECAM $VR$-band data with a longer exposure than the $g$-band image. 
The radio contours are shown in white for clarity. North is up and east to the left.}\label{fig:1}
\end{figure*}

\section{Association with 4FGL~J1647.5$-$5724}\label{sec2}
\subsection{Updating the Gamma-ray Localization}
The optimized \gm-ray position of 4FGL~J1647.5$-$5724 reported in the 4FGL-DR2 catalog has not been updated in the subsequent data releases, 4FGL-DR3, and 4FGL-DR4. Since the 4FGL-DR2 catalog was prepared using ten years of the Fermi-LAT data, we analyzed the LAT data covering $\sim$15.7 years (2008 August 4$-$2024 April 5) of the Fermi operation to update the \gm-ray position. In the energy range of 0.1$-$300 GeV, the standard data reduction procedure was adopted, harnessing the full potential of the Pass8 dataset by considering the division of photons by point-spread function (PSF) event types \citep[see, e.g.,][for details]{2022ApJS..260...53A}. The estimated best-fitted \gm-ray position and 95\% uncertainty radius ($R_{\rm 95}$) are right ascension (RA) = 251$^{\circ}$.884 and declination (Dec) = $-57^{\circ}$.450 and 0$^{\circ}$.068, respectively. With the addition of the larger dataset, hence deeper sensitivity, the optimized \gm-ray position shifted closer towards the collisional ring galaxy system, thus suggesting a potential association (Figure~\ref{fig:1}, top left panel). The derived test statistic is 53.7, compared to 43.9 reported in the 4FGL-DR2 catalog. The power law photon flux and index values derived from the LAT data analysis are $(1.01\pm0.25)\times10^{-8}$ \phflux~and $2.51\pm0.10$, respectively. At 10 Mpc distance, this corresponds to an isotropic luminosity of $(4.50\pm0.79)\times10^{40}$ \lum. The \gm-ray spectral shape appears to be softer compared to that observed from star-forming galaxies but consistent within errors \citep[average photon index = $2.18\pm0.29$,][]{2020ApJ...894...88A}.

{\it ``Kathryn's Wheel"} is the nearest collisional ring galaxy system and lies within the 95\% positional uncertainty region of 4FGL~J1647.5$-$5724. Located in the low-latitude Galactic neighborhood ($|b|=-8^{\circ}$, distance $\sim$10 Mpc), the system includes a dwarf irregular galaxy, LEDA~3080069, acting as the `bullet' that has collided with an edge-on, late-type, spiral galaxy ESO~179$-$13 \citep[e.g.,][]{1987ess..book.....L,2015MNRAS.452.3759P}. The system also contains another dwarf irregular galaxy LEDA~3080065 (Figure~\ref{fig:1}, top left panel). The system has an overall physical size of $\sim$15 kpc, a total mass of $6.7\times10^9$ \msun~(H{\sc i} + stars), and a metallicity of [O/H]$\sim-0.4$, thus indicating it to be a Magellanic-type system \citep[][]{2015MNRAS.452.3759P}. To look for possible blazars lying within $R_{\rm 95}$, we used the BZCAT, Wide-field Infrared Survey Explorer (WISE) Blazar-like Radio-Loud Sources, and Kernel Density Estimation-selected candidate BL Lacs catalogs \citep[][]{2015Ap&SS.357...75M,2019ApJS..242....4D}. No known blazars or radio galaxies are found within the \gm-ray uncertainty region, thus strengthening the association of the \gm-ray object with {\it ``Kathryn's Wheel"}.

\subsection{X-ray Observations}
 We analyzed 13.1 ksec Swift X-Ray Telescope (XRT) data to identify potential X-ray counterparts within $R_{\rm 95}$ using the Swift-XRT online data products generator\footnote{\url{https://www.swift.ac.uk/user\_objects/}} \citep[][]{2009MNRAS.397.1177E}. The details of the source finding algorithm adopted in this tool are given in \citet[][]{2020ApJS..247...54E}. The generated counts map is shown in the top right panel of Figure~\ref{fig:1}. Three X-ray point sources were found, with two of them being located at $\sim$13$^{\prime\prime}$ eastwards and $\sim$33$^{\prime\prime}$ westwards of the dwarf galaxy LEDA 3080069. The third and brightest X-ray source is positionally consistent with the 7.7 mag A0IV star HD~150915. Given the faintness of X-ray sources located close to the dwarf galaxy, it is not possible to constrain their spectral parameters. Deeper X-ray observations will be needed to ascertain their physical properties. 

\begin{figure*}
    \hbox{\hspace{1.7cm}
    \includegraphics[scale=0.35]{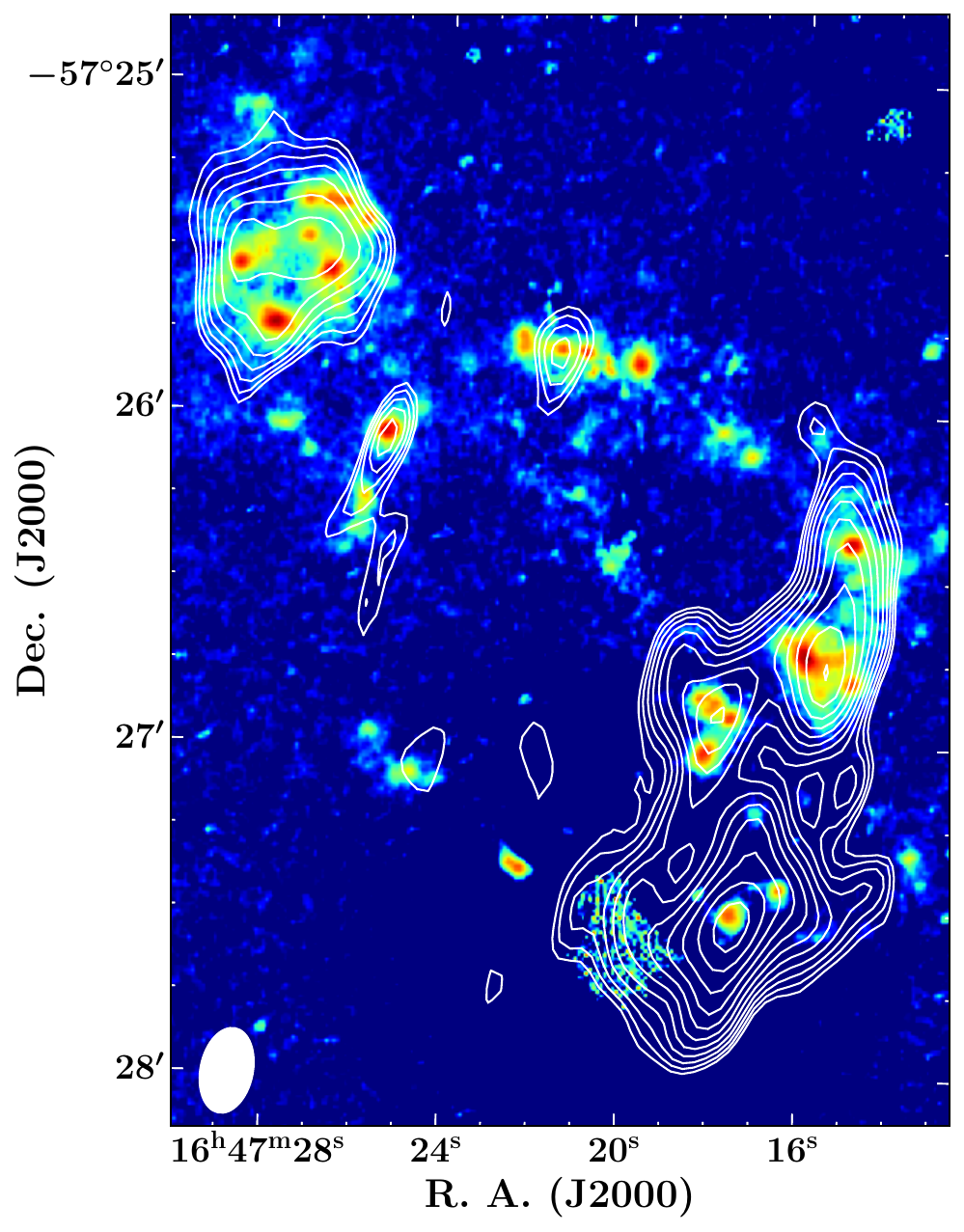}
    \includegraphics[scale=0.45]{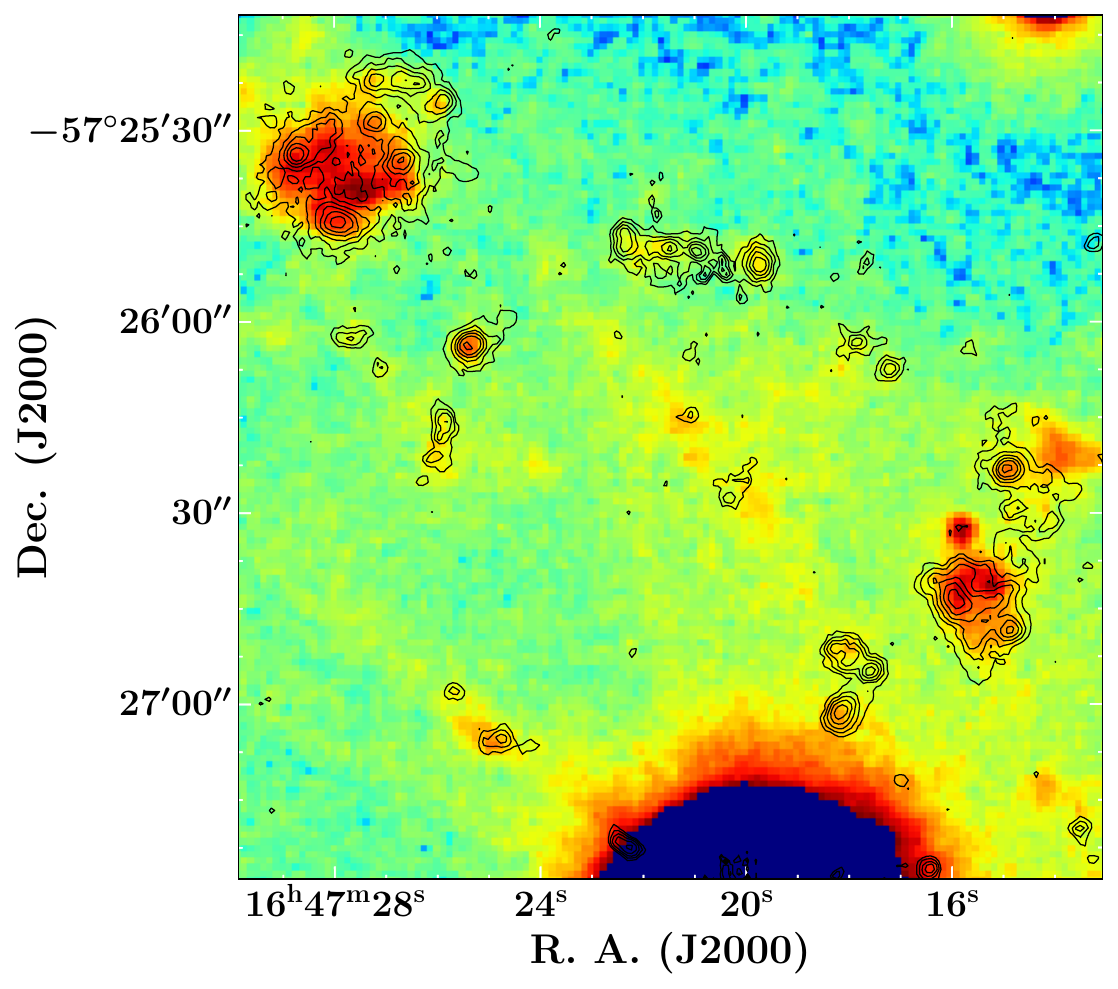}
    }
\caption{Left: The continuum-subtracted H$\alpha$ image of the collisional ring galaxy system. We overplot the 1.36 GHz RACS contours to highlight the co-spatiality of the star-forming regions and the radio emission. Right: The $W2$-band image of the system taken with Swift-UVOT. The overplotted H$\alpha$ contours (representing the star-forming regions shown in the left panel) reveal that the co-spatial bright UV emission likely originated from the massive hot stars. North is up and east to the left.}\label{fig:2}
\end{figure*}

\subsection{Radio Observations}
We checked radio sources located within $R_{\rm 95}$ using the second data release of the Rapid ASKAP Continuum Survey \citep[RACS-mid;][]{2024PASA...41....3D}. At the central wavelength of 1.36 GHz, this Survey covers the whole sky south of +49$^{\circ}$ declination and has a median angular resolution of 11$^{\prime\prime}$.2$\times$9$^{\prime\prime}$.3 though it varies to maximize sky coverage and sensitivity across the covered area. The cutout image covering the region of interest was downloaded from the RACS data server\footnote{\url{https://research.csiro.au/casda/}} and overplotted on the Swift-XRT and the $VR$-band optical image taken with the Dark Energy Camera \citep[DECAM;][]{2021ApJS..255...20A}. A complex, diffuse radio morphology was seen overlapping with {\it ``Kathryn's Wheel"} though the central galaxy, ESO 179$-$13, remain undetected down to 0.45 mJy/beam, i.e., 3$\sigma$ detection limit (Figure~\ref{fig:1}, bottom panel). None of the identified X-ray sources exhibit a radio morphology typically observed in an AGN, thus ruling out the possibility of them being jetted AGN \citep[e.g.,][]{2011ApJ...741...30A}. A bright radio object ($F_{\nu,\rm 1.36GHz}\sim67$ mJy) was found at 4.7$^{\prime}$ eastwards from LEDA~3080069 (Figure~\ref{fig:1}, top right panel). This object is unlikely to be the blazar counterpart of 4FGL~J1647.5$-$5724 because: (i) it is outside of the updated $R_{\rm 95}$, (ii) it remained undetected in Swift-XRT observations, (iii) its optical counterpart (RA=252$^{\circ}$.004979, Dec=$-$57$^{\circ}$.405146) shows a large proper motion of 9.1 mas year$^{-1}$ in Gaia observations \citep[][]{2021A&A...649A...1G} indicating a possible Galactic origin, and (iv) in the WISE color-color diagram, it occupies a region far from that is populated by \gm-ray emitting blazars \citep[][]{2012ApJ...750..138M}.

A detailed morphological study of {\it ``Kathryn's Wheel"} was carried out by \citet[][]{2015MNRAS.452.3759P}. The source was observed as a part of the SuperCOSMOS H${\alpha}$ Survey of the Southern Galactic plane \citep[][]{2005MNRAS.362..689P} which revealed the central galaxy, ESO 179$-$13, to be surrounded by a ring of star-forming knots and having 6.1 kpc diameter. Based on the H${\alpha}$ and mid-infrared flux measurements, the integrated star-formation rate (SFR) of the whole system is reported to be 0.2$-$0.5 M$_{\odot}$ year$^{-1}$, higher than that of Large Magellanic Cloud \citep[LMC;][]{2015MNRAS.452.3759P}. The most luminous H{\sc ii} region in the system has the H$\alpha$ luminosity, $\simeq 5\times10^{39}$ \lum, comparable to supergiant H{\sc ii} region 30 Doradus in LMC, which is also a bright \gm-ray emitter \citep[][]{2010A&A...512A...7A}. These results provide supportive evidence that the vigorous star-forming activities of {\it ``Kathryn's Wheel"} could be responsible for the observed \gm-ray emission, thereby making it a plausible counterpart of the \gm-ray source 4FGL~J1647.5$-$5724.

\section{Star-forming Ring, Bullet Galaxy, and the Gamma-ray Emission}\label{sec3}
The H$\alpha$ and the matching broad-band short-red images of {\it ``Kathryn's Wheel"} were obtained from the SuperCOSMOS H$\alpha$ survey website\footnote{\url{http://www-wfau.roe.ac.uk/sss/halpha/hapixel.html}}. We show the continuum-subtracted image, i.e., taking out the short-red image from the H$\alpha$ map, in the left panel of Figure~\ref{fig:2} to highlight the zones of star formation. The spectacular star-forming ring surrounding the central galaxy, ESO 179$-$13, is evident, which was also reported by \citet[][]{2015MNRAS.452.3759P}. The central galaxy shows little star formation, possibly stripped of gas due to interaction with the `bullet' galaxy LEDA~3080069. This is further illustrated in the right panel of Figure~\ref{fig:2} where we have shown the $W2$-filter ($\lambda_{\rm central}=1928$ \AA) image of the region taken with the Swift Ultraviolet and Optical Telescope. The star-forming regions dominated by the massive, hot stars, thus brighter in the ultraviolet band, are found to be coincident with the H$\alpha$ emitting ring. We overplotted the 1.36 GHz RACS data on the continuum-subtracted H$\alpha$ image to explore the possible connection of the observed radio emission with the star-forming regions (Figure~\ref{fig:2}, left panel). The radio contours were found to overlap with the H$\alpha$ emitting knots in the ring and the dwarf galaxies. These observations suggest a close connection of the radio emission with the star formation and its origin to be the diffuse synchrotron radiation produced by cosmic-ray particles spiraling in the interstellar medium magnetic fields \citep[e.g.,][]{1992ARA&A..30..575C}. Considering integrated flux densities of the system measured at 887.5 MHz and 1.36 GHz, the radio spectral index is $\alpha=-2.1$ ($F_{\nu}\propto \nu^{\alpha}$). Such a steep spectrum further supports the synchrotron nature of the observed radio emission.

\begin{figure}
\hbox{
    \includegraphics[width=\linewidth]{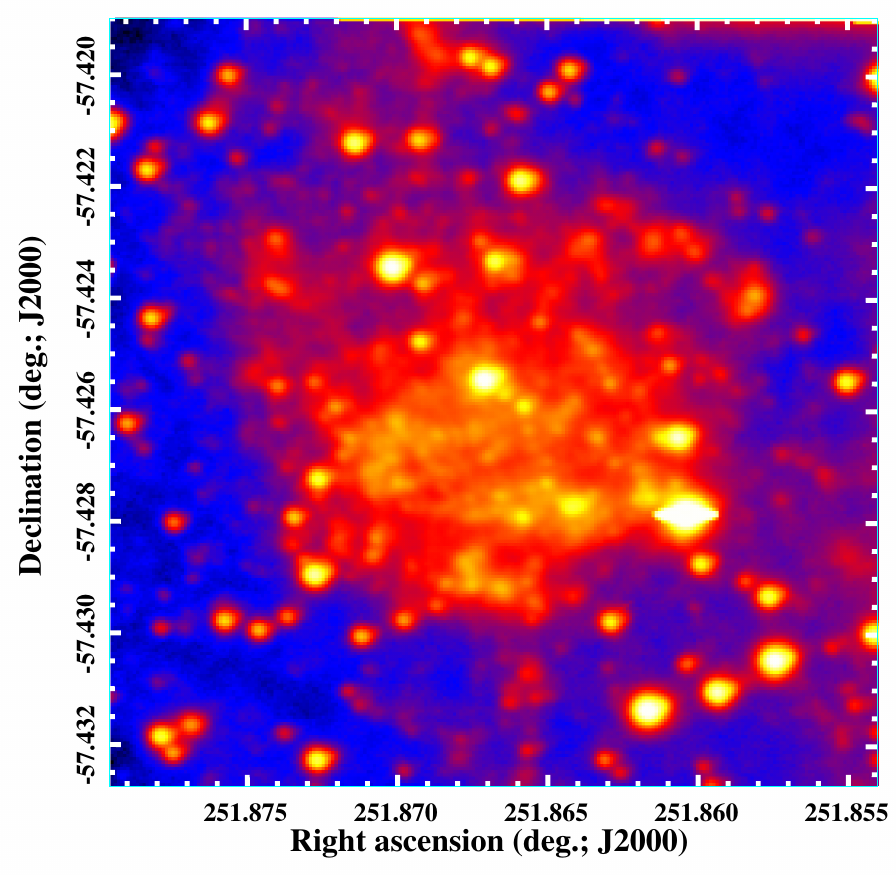}
    }
\caption{The 0$^{\prime}$.8$\times$0$^{\prime}$.8 VIMOS-VLT image of the dwarf galaxy LEDA 3080069 centered at RA=251$^{\circ}$.8667055, and Dec=$-$57$^{\circ}$.4258147 (J2000). North is up and east to the left.}\label{fig:bullet}
\end{figure}

The continuum subtracted H$\alpha$ image has revealed that the `bullet' galaxy, LEDA~3080069, is undergoing intense star formation, which is likely triggered due to collision with ESO~179$-$13 \citep[Figure~\ref{fig:2}, see also,][]{2015MNRAS.452.3759P}. In Figure~\ref{fig:bullet}, we show its high-resolution $V$-band image taken with the VIsible MultiObject Spectrograph (VIMOS, now decommissioned) at Very Large Telescope (VLT, proposal id: 099.B-0215, PI: A. A. Zijlstra). Several patchy star-forming regions are visible, and the overall morphology is consistent, with this object being a dwarf irregular galaxy.

The massive hot stars in the star-forming regions radiate primarily at ultraviolet wavelengths, which is reprocessed by interstellar dust to the infrared band. Therefore, IR emission can be considered as a proxy for SFR \citep[e.g.,][]{1998ARA&A..36..189K}. These stars usually end their lives as core-collapse supernovae, and the produced shocks can efficiently accelerate cosmic rays. The interaction of cosmic rays with interstellar gas and interstellar radiation fields via bremsstrahlung and inverse Compton scattering processes, respectively, is thought to be responsible for diffuse \gm-ray emission observed from star-forming galaxies. Therefore, a strong correlation between the \gm-ray and infrared luminosities is expected and has been reported \citep[cf.][]{2012ApJ...755..164A,2020ApJ...894...88A,2020A&A...641A.147K}. Since the integrated infrared luminosity, $L_{\rm 8-1000\mu m}$ is unknown for {\it ``Kathryn's Wheel"}, we adopted the following conversion proposed by \citet[][]{1998ApJ...498..541K}:

\begin{equation}
\frac{\rm SFR}{{\rm M}_{\odot}\:{\rm year}^{-1}}=\Psi 1.7 \times 10^{-10} \frac{L_{8-1000
\; \mu{\rm m}}}{L_{\odot}},
\label{eq_sfr_ir}
\end{equation}
where the factor $\Psi$ depends on the assumed initial mass function (IMF) and taken as $\Psi=0.79$ considering \citet[][]{2003PASP..115..763C} IMF. \citet[][]{2015MNRAS.452.3759P} reported the SFR for the system to be in the range 0.2$-$0.5 \msun~year$^{-1}$ which suggests the $L_{\rm 8-1000\mu m}$ to be $(1.5-3.7)\times10^9$ \lsun. 

\begin{figure}
\hbox{
    \includegraphics[width=\linewidth]{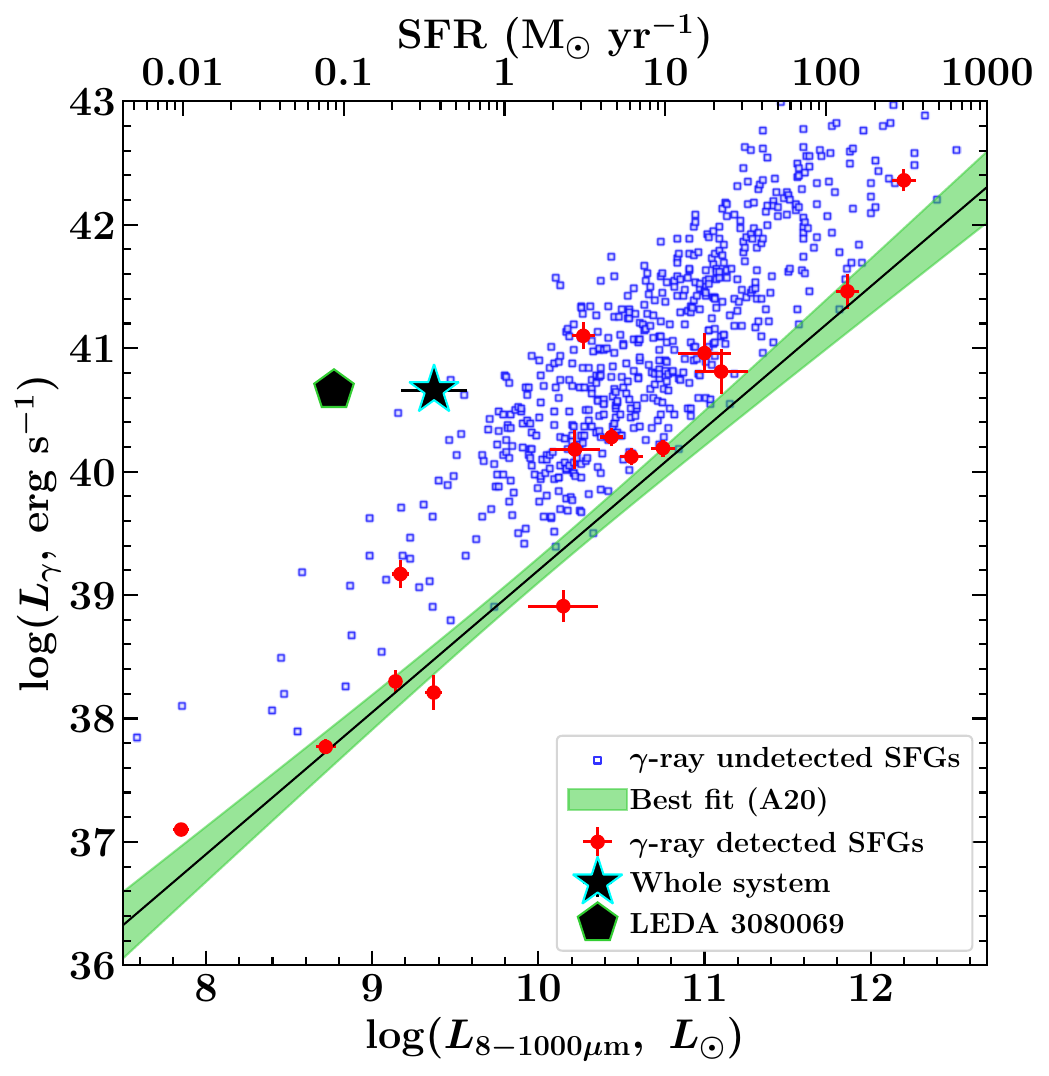}
    }
\caption{This plot shows the \gm-ray luminosity as a function of infrared luminosity for \gm-ray detected (red circles) and undetected (blue squares) star-forming galaxies. For Fermi-LAT undetected sources, the \gm-ray luminosities are upper limits. The black solid line and green shaded region refer to the best-fitted correlation and 1$\sigma$ uncertainty, respectively, reported by \citet[][]{2020ApJ...894...88A}. The location of {\it ``Kathryn's Wheel"} system is shown with a black star. The black pentagon shows the case when all of the \gm-ray emission is considered to be originating from the dwarf galaxy LEDA 3080069 alone.}\label{fig:corr}
\end{figure}

Figure~\ref{fig:corr} shows the relation between the \gm-ray and infrared luminosities for \gm-ray detected and undetected star-forming galaxies studied by \citet[][]{2020ApJ...894...88A}. We also include {\it ``Kathryn's Wheel"} in this diagram, which appears to lie above the best-fitted correlation line, i.e., it has a \gm-ray luminosity larger than that predicted by the correlation. One possible explanation for this discrepancy is that the SFR measured using H$\alpha$ luminosity or mid-infrared flux could be affected due to strong contamination from the nearby star HD~150915. An over-subtraction of the stellar emission may underestimate the H$\alpha$ luminosity, thus SFR, and, in turn, $L_{\rm 8-1000\mu m}$. If the intrinsic SFR is higher, the system's location will shift closer to the best-fit correlation line. Furthermore, given the proximity of {\it ``Kathryn's Wheel"} to the Milky Way, there could be a \gm-ray emitting population of stellar remnants, e.g., pulsars and pulsar wind nebulae, residing in the system that might be contributing to the observed \gm-ray emission. Indeed, the \gm-ray emitting pulsars, J0540$-$6919 and J0537$–$6910, found in the 30 Doradus nebula of LMC contribute $\sim$60\% and $\sim$25\% of the observed GeV flux density from this region, respectively, indicating only a minor fraction to be produced by cosmic rays \citep[][]{2015Sci...350..801F}. These results also rule out the possibility of the \gm-ray emission originating from the star-forming regions in the `bullet' galaxy LEDA~3080069 alone. This is because it has $\sim$24\% contribution to the total SFR \citep[][]{2015MNRAS.452.3759P} which would push the source even farther from the best-fit correlation (Figure~\ref{fig:corr}).

Finally, galaxy collisions/mergers are known to trigger not only the star formation but also the AGN phenomenon \citep[][]{2006ApJS..163....1H,2020A&A...637A..94G}. Therefore, it is possible that there could be AGN activity buried in dwarf galaxies of the system, i.e., LEDA 3080065 and LEDA 3080069, that may account for some of the observed \gm-ray emission. Indeed, star-forming galaxies, e.g., NGC 3424 and UGC 11041, show significant offset from the best-fit correlation shown in Figure~\ref{fig:corr}. A non-negligible contribution from the AGN activities in these objects has not been discarded \citep[][]{2019ApJ...884...91P}. Interestingly, the \gm-ray emission observed from {\it ``Kathryn's Wheel"} is non-variable, and the non-detection of a radio core-jet morphology disfavors the presence of a relativistic jet. Arcsecond/sub-arcsecond-scale radio and X-ray observations will be needed to examine these hypotheses. All in all, {\it ``Kathryn's Wheel"} can be considered a test-bed to explore the origin and transport of cosmic rays and their connection with star-forming activities given its proximity to the Milky Way.

\acknowledgements
We thank the journal referee for constructive criticism. Based on observations made with ESO Telescopes at the La Silla Paranal Observatory under programme ID 099.B-0215. This work made use of data supplied by the UK Swift Science Data Centre at the University of Leicester. This paper includes archived data obtained through the CSIRO ASKAP Science Data Archive, CASDA. This research uses services or data provided by the Astro Data Lab, which is part of the Community Science and Data Center (CSDC) Program of NSF NOIRLab.

\bibliographystyle{aasjournal}
\bibliography{Master}

\end{document}